\documentclass[11pt,a4paper]{article}
\pdfoutput=1
\usepackage{jheppub}

\begin{document}

\title{Spin-one dark matter and gamma ray signals from the galactic center.}
\author[a]{H. Hern\'{a}ndez-Arellano,}
\author[a]{M. Napsuciale}
\author[b]{and S. Rodr\'{\i}guez.}
\affiliation[a]{Departamento de F\'{i}sica, Universidad de Guanajuato, Lomas del Campestre 103, Fraccionamiento
Lomas del Campestre, Le\'on, Guanajuato M\'exico, 37150.}
\affiliation[b]{Facultad de Ciencias F\'isico-Matem\'aticas,
  Universidad Aut\'onoma de Coahuila, Edificio A, Unidad
  Camporredondo, 25000, Saltillo Coahuila M\'exico.} 
\emailAdd{h.hernandezarellano@ugto.mx}
\emailAdd{mauro@fisica.ugto.mx}
\emailAdd{simonrodriguez@uadec.edu.mx}

\begin{abstract}
{ In this work we study the possibility that the gamma ray excess (GRE) at the Milky Way galactic center come from
the  annihilation of dark matter with a $(1,0)\oplus(0,1)$ space-time structure (spin-one dark matter, SODM). 
We calculate the production of prompt photons from initial state radiation, internal bremsstrahlung, final state radiation including  the 
emission from  the decay products of the $\mu, \tau$  or hadronization of quarks.  Next we study the delayed photon emission from the 
inverse Compton scattering (ICS) of electrons (produced directly or in the prompt decay of  $\mu, \tau$ leptons or 
in the hadronization of quarks produced in the annihilation of SODM) with the cosmic microwave background or starlight. All these 
mechanisms yield significant contributions only for Higgs resonant exchange, i.e. for $M\approx M_{H}/2$, and the results depend 
on the Higgs scalar coupling to SODM, $g_{s}$. The dominant mechanism at the GRE bump is the prompt photon production in the 
hadronization of $b$ quarks produced in $\bar{D}D\to \bar{b}b$, whereas the delayed photon emission from the ICS of electrons coming 
from the hadronization of $b$ quarks produced in the same reaction 
dominates at low energies ($\omega < 0.3~ GeV$) and prompt photons from $c$ and $\tau$, as well as from internal bremsstrahlung, 
yield competitive contributions at the end point of the spectrum ($\omega \ge 30 ~GeV$). 

Taking into account all these contributions, our results for photons produced in the annihilation of SODM are in good agreement with the 
GRE data for $g_{s}\in [0.98, 1.01] \times 10^{-3}$ and $M\in [62.470,62.505]~GeV$. 
We study the consistency of the corresponding results for the dark matter relic density,  the spin-independent dark matter-nucleon 
cross-section  $\sigma_{p}$ and the cross section for the annihilation of dark matter into 
$\bar{b}b$, $\tau^{+}\tau^{-}$, $\mu^{+}\mu^{-}$ and $\gamma\gamma$, taking into account the Higgs resonance effects, 
finding consistent results in all cases.
}
\end{abstract}
\maketitle

\section{Introduction}
The understanding of the nature of dark matter is presently one of the major challenges in particle physics, astrophysics and cosmology. 
Dark matter amounts for $26\%$ of the content of the universe and there is compelling evidence for its existence from the 
measurements of several independent observables, among which we have galaxy rotation curves, cosmic microwave background and
dark matter relic density (for a recent review with a comprehensive list of references see \cite{Lin:2019uvt}).

 The measured dark matter relic density \cite{Tanabashi:2018oca} can be obtained from its thermal 
 decoupling from the primordial plasma. This requires $\langle \sigma v_{r} \rangle \approx 10^{-9} GeV^{-2}\approx 10^{-26} cm^{3}/seg$, 
 i.e. of the order of the weak scale cross sections, and a mass of the order of a hundred GeV for the so called weakly interactive 
 massive particle (WIMP). 
 This result can be understood in terms of the exchange of a massive particle between dark matter and standard model (SM) 
 fields, with couplings of the order of the weak interactions, which points to the unification route, i.e. to identify dark matter with particles 
 arising in formalisms unifying the three interactions of the SM or including gravity with a wide variety of models yielding candidates 
 with different  space-time structures.  Recently, a thorough study of the possibilities for scalar, fermionic and vector dark matter was 
 done in \cite{Arcadi:2017kky}, concluding that little room is left  by available data for WIMPs with these space-time structures 
 (see however \cite{Carena:2019pwq}). 
 
 The SM  uses only a  very restricted set of the Homogeneous Lorentz Group (HLG) irreducible representations (irreps). Indeed, 
from the isomorphism of the HLG with $SU(2)\otimes SU(2)$, the irreps of the HLG can be labelled by two $SU(2)$ quantum numbers 
$(a,b)$.  The SM  uses only the scalar irrep, $(0,0)$, for the Higgs, the spin $1/2$ chiral representations $(\frac{1}{2},0)$ and 
$(0,\frac{1}{2})$, for quarks and leptons and the vector irrep, $(\frac{1}{2},\frac{1}{2})$, for the gauge fields. 
Quantum field theory proposals for physics beyond the SM in general use these very same
representations for their field content, except for supergravity which includes the gravitino transforming in the 
$(1,\frac{1}{2})\oplus (\frac{1}{2},1)$ representation and the graviton transforming in the $(1,1)$ representation.
 
 In Ref. \cite{Hernandez-Arellano:2018sen} we  proposed an alternative  $(1,0)\oplus (0,1)$ space-time structure for dark matter,  
 in a formalism that generalizes the structure of spin 1/2 Dirac theory to spin-one matter particles. The corresponding quantum field theory 
 was developed in \cite{Napsuciale:2015kua} and it is based in the parity-based construction of a covariant basis for 
 $(j,0)\oplus (0,j)$ fields done in \cite{Gomez-Avila:2013qaa}. Spin-one matter fields are described by a six-component 
 spinor and can be endowed with a vector gauge 
 structure (the kinetic term is not chiral thus chiral gauge interactions are not permitted). The interaction of spin-one dark matter (SODM) 
 with SM fields is constructed under the effective field theory philosophy and a basic principle: SM fields are singlets of the 
 dark gauge group and viceversa. Considering for simplicity the dark gauge group as $U(1)_{D}$, the leading interacting terms in the 
 effective field theory are   \cite{Hernandez-Arellano:2018sen} 
 \begin{equation}
\mathcal{L}_{int}=\bar{\psi}(g_{s}\mathbf{1}+ig_{p}\chi )\psi \tilde{\phi}%
\phi +g_{t}\bar{\psi}M_{\mu \nu }\psi B^{\mu \nu } + \mathcal{L}_{selfint},  
\label{Leff}
\end{equation}%
where $g_{s}$, $g_{p}$ and $g_{t}$ are low energy constants. The self-interaction terms $ \mathcal{L}_{selfint}$ are given 
in \cite{Napsuciale:2015kua} and are not relevant for the purposes of this work. After spontaneous symmetry breaking, 
this Lagrangian yields 
\begin{equation}
\mathcal{L}_{int}= \frac{1}{2}\bar\psi (g_{s} \mathbf{1}+ i g_{p}\chi) \psi
\left(H+ v\right)^2 + g_{t} C_{W} \bar\psi M_{\mu\nu} \psi
F^{\mu\nu} - g_{t} S_{W} \bar\psi M_{\mu\nu} \psi Z^{\mu\nu} ,
\label{lag}
\end{equation}%
where $C_W$, $S_W$, $H$, $v$, $F^{\mu\nu}, Z^{\mu\nu}$ are the cosine and sine of the Weinberg angle, the Higgs field, 
the Higgs vacuum expectation value and the electromagnetic and $%
Z^0$ stress tensors, respectively. The interacting terms include a spin portal (photon and $Z^{0}$ coupling to higher multipoles 
of dark matter), a Higgs portal with a scalar and a parity-violating pseudo-scalar interaction, and a dark matter - anti-dark matter 
to two Higgs interaction.

Light dark matter ($M<M_{Z}/2$) turns out to be inconsistent with the measured relic density and the generated invisible widths of 
the $Z^{0}$ and Higgs boson. The upper bounds for the termal average 
$\langle \sigma v_{r} \rangle$ for $\bar{D}D \to \bar{b}b, \tau^{+}\tau^{-}$ extracted by the FermiLAT-DES collaboration in the analysis of  
photon signals coming from Milky Way dwarf satellite galaxies 
\cite{Drlica-Wagner:2015xua} \cite{ Fermi-LAT:2016uux} are well satisfied. On the direct detection side, the appropriate description 
of XENON1T measurement of $\sigma_{p}$ \cite{Aprile:2017iyp} requires 
spin portal coupling $g_{t} \leq 10^{-5}$ for $M\approx 100~GeV$ to   $g_{t} \leq 10^{-3}$ for $M\approx 1~TeV$. Less stringent 
upper bounds are obtained for the Higgs portal coupling $g_{s}\leq 10^{-2}$ and no significant constraint is obtained for $g_{p}$ 
\cite{Hernandez-Arellano:2018sen}.

In this formalism, the annihilation of spin-one dark matter into final states containing photons occurs at tree level which can be relevant 
for the search of gamma rays coming from annihilation of dark matter. Gamma ray searches started a new era with the launch of  
PAMELA \cite{Orsi:2007zz},  AMS-02 \cite{Battiston:2008zza}, and the Large Area Telescope of the Fermi experiment (FermiLAT) \cite{Moiseev:2008zz}. Presently, strong upper limits have been obtained by these collaborations for the annihilation of dark matter 
into SM particles, which are complementary to searches at higher energies in experiments like HESS \cite{Abdallah:2018qtu}. 

In the present work we study photon signals from the annihilation of SODM and the possible contributions of the corresponding mechanism 
to the gamma ray excess (GRE) from the galactic center claimed by several collaborations.  The next section is devoted to the definition of 
the observables and conventions in this work and to the the key observation that SODM yield annihilation cross sections of the order of 
the expected by recent fit to the GRE only at the Higgs resonance. 
In Section III we study the mechanisms for the production of prompt photons in the annihilation of SODM. In Section IV we work out the 
production of gamma rays from inverse Compton scattering of electrons produced directly or indirectly in the annihilation of SODM. 
In Section V we collect and give our final results for this study. The consistency of our results with existing constraints 
on several observables is analyzed in Section VI. Finally, our conclusions are given in Section VII.

\section{Gamma ray excess from the galactic center}
 \subsection{GRE, prompt photons and muon inverse Compton scattering}
During the past few years, it has been claimed by several groups that an excess over the expected gamma ray flux from known 
sources in the Milky Way galactic center exists in the FermiLAT data  around $3~GeV$ 
\cite{Hooper:2010mq}\cite{Boyarsky:2010dr}\cite{Hooper:2011ti}\cite{Abazajian:2012pn}
 \cite{Macias:2013vya}\cite{Gordon:2013vta}\cite{Abazajian:2014fta}\cite{Daylan:2014rsa}\cite{Calore:2014xka}\cite{Zhou:2014lva}\cite{TheFermi-LAT:2015kwa}. The large uncertainties involved in the interpretation of FermiLAT data have been recently analyzed 
 by the FermiLAT collaboration \cite{TheFermi-LAT:2017vmf} concluding that a GRE excess 
 in a region around 3 GeV indeed exists, but a  broad band of possible values for the corresponding differential flux as a function of the 
 photon energy is permitted by these uncertainties. Although the GRE can be explained by little known astrophysical sources 
 \cite{Hooper:2010mq} \cite{YusefZadeh:2012nh} \cite{Linden:2012iv} \cite{Carlson:2014cwa} \cite{Petrovic:2014uda} 
 \cite{Cholis:2014lta}, the annihilation of dark matter into final states containing photons remains as an attractive possibility 
 \cite{Carena:2019pwq} \cite{Cline:2013gha} \cite{Okada:2013bna} \cite{Abazajian:2014fta} \cite{Basak:2014sza} \cite{ Agrawal:2014oha}  
  \cite{Lacroix:2014eea} \cite{Calore:2014nla}   
\cite{Duerr:2015bea} \cite{ Cuoco:2016jqt} \cite{ Sage:2016xkb}   and we work out here the results for 
dark matter with a $(1,0)\oplus (0,1)$ space-time structure.   

The gamma ray differential flux from the annihilation of (non-self-conjugated) dark matter is
\begin{equation}
 \frac{d\Phi}{d\omega} =\left( \sum_i \frac{B_i}{16 \pi M^2} \frac{d \langle \sigma v_{r} \rangle^{\gamma}_i}{d\omega} \right)
  \int_{\Delta \Omega} \int_{l.o.s} \rho^2 (r(s,\theta)) ds d\Omega .
  \label{flux}
\end{equation}
 The sum runs over all annihilation channels containing at least one photon in the final state; 
 $\frac{d \langle \sigma v_{r} \rangle^{\gamma}_i}{d\omega}$ is the velocity averaged differential cross section for the $i$-channel 
 and $B_i$ is the number of photons produced in this process. In the literature, the differential photon flux in Eq.(\ref{flux}) is sometimes 
 written in terms of the spectrum for each channel
\begin{equation}
\frac{dN^{\gamma}_{i}}{d\omega}\equiv\frac{1}{\langle \sigma v_{r} \rangle_{i} }  \frac{d \langle \sigma v_{r} \rangle^{i}_{\gamma} }{d\omega} ,
\end{equation}
where $\langle \sigma v_{r} \rangle_{i}$ stands for the non-radiative cross section for the $i$-channel. It is conventionally assumed 
that this non-radiative cross section contains all the information on the annihilation of dark matter entering the radiative process, 
in such a way that the spectrum has the information of the photon production from standard model $i$-states which can be 
calculated and, beyond technical details, it is well known. In this construction, model independent fits to data can be done 
with $\langle \sigma v_{r} \rangle_{i}$ and $M$ as free parameters. 

 The term in the parentheses  in Eq.(\ref{flux}) contains all the information from the dark matter interactions and the integral contains 
 the so called \textit{J-factor}  for the observation window defined by the solid angle $\Delta\Omega $
 \begin{equation}
 J(\Delta\Omega )=  \int_{\Delta \Omega} \int_{l.o.s} \rho^2 (r(s,\theta)) ds d\Omega.
 \end{equation}
  In the computations in this work, unless otherwise explicitly stated, we will use the generalized Navarro-Frenk-White (gNFW) dark matter 
  profile
 \begin{equation}
 \rho (r)=\rho_{s} \frac{r^{3}_{s} }{r^{\gamma} (r+r_{s})^{3-\gamma}  } 
 \end{equation}
 with the values $r_{s}=20 ~kpc $ for the scale radius, $\gamma=1.25$ for the slope of the inner part of the profile. The scale 
 density $\rho_{s}=0.225~GeV/cm^{3}$ is fixed by requiring the local dark matter density at $r_{\odot}=8.5~ kpc$ to be 
 $\rho_{\odot}=0.4~ GeV/cm^{3}$. 
 
The production of prompt gamma rays in the annihilation of dark matter into 
$\bar{q}q,\bar{c}c,\bar{b}b$, $e^{+}e^{-},\mu^{+}\mu^{-},\tau^{+}\tau^{-}, W^{+}W^{-}, ZZ, hh, gg$  was fitted to the FermiLAT data for the 
GRE in \cite{Calore:2014nla},  finding in general that the GRE can be explained if dark matter annihilates into any 
of these pairs of particles  except for the $e^{+}e^{-}$ channel, whenever the dark matter mass is in the range $5-174 ~GeV$ depending on the 
specific channel, and the corresponding cross section is of the order of the thermal one, 
$\langle \sigma v_{r} \rangle\approx 10^{-26} cm^{3}/seg$.  In particular, the annihilation into fermionic states, $\bar{D}D\to \bar{f}f$ 
with  $f=\mu,\tau,q,c,b$ yields good fits to the GRE data for dark matter mass in the $9-61~ GeV$ range. 

Another possible mechanism to explain the GRE is the production of gamma rays in the Inverse Compton Scattering (ICS) of electrons 
and muons produced in dark matter annihilation into $e^{+}e^{-}$ and $\mu^{+}\mu^{-}$, which propagates over the galactic center and 
scatter photons from the Cosmic Microwave Background (CMB) or starlight \cite{Lacroix:2014eea}. The electronic channel for electrons 
produced directly in the annihilation of dark matter into $e^{+}e^{-}$,  requires a large 
annihilation cross section $\langle \sigma v_{r} \rangle_{e}$ which is severely constrained by the positron fraction data from the 
AMS Collaboration \cite{Bergstrom:2013jra} \cite{Aguilar:2013qda}. In \cite{Calore:2014nla}, it was shown that the muon channel 
yield sizable contributions to the GRE and when added to the prompt photon production in this channel allows for higher values of the 
dark matter mass ($M\approx 61~GeV$), with the required cross section still being of the order of the thermal cross section. 

 \subsection{Spin-one dark matter annihilation into fermions and the GRE}

The interactions of SODM with SM fields in Eq. (\ref{lag}) allow for the annihilation of dark matter into all the channels mentioned above 
except for the $gg$ channel.  The cross-section for the annihilation of SODM into a fermion pair, worked out 
in \cite{Hernandez-Arellano:2018sen}, is
\begin{align}
(\sigma v_r)_{\bar{f}f} (s)& =\frac{1}{ 144\pi M^4 \sqrt{s}} \frac{\sqrt{%
s-4m_{f}^{2}}}{(s-M^2)} \left[ \frac{ m_f^2 \left(s-4 m_f^2\right) \left( g_p^2 s
\left(s-4 M^2\right)+g_s^2 \left(6 M^4-4 M^2 s+s^2\right)\right)}{
\left(\left(s-M_H^2\right)^2 + \Gamma_H^2 M_H^2\right)} \right.  \notag \\
&\left. +\frac{2 g_{t}^2 M_Z^2 S_W^2 s\left(s-4 M^2 \right) \left(2
M^2+s\right) \left(2 \left(A_{f}^2-2 B_{f}^2\right) m_f^2+s
\left(A_{f}^2+B_{f}^2\right)\right)}{3 v^2 \left(\left(s-M_Z^2\right)^2 + \Gamma_Z^2 M_Z^2\right){}^2} \right.  \notag
\\
&\left. +\frac{32 C_W^2 Q_f^2 g_{t}^2 M_W^2 S_W^2 \left(s-4 M^2 \right)
\left(2 M^2+s\right) \left(2 m_f^2+s\right)}{3 v^2 s} \right.  \notag \\
&\left. -\frac{16 A_{f} C_W Q_f g_{t}^2 M_W M_Z S_W^2 \left(s-4 M^2 \right)
\left(2 M^2+s\right) \left(2 m_f^2+s\right)}{3 v^2 \left(\left(s-M_Z^2\right)^2 + \Gamma_Z^2 M_Z^2\right)} %
\right].
\label{sigmavffcomp}
\end{align}
where $m_f$, $Q_f$ correspond to the mass of the fermion and its charge in units of $e>0$, respectively, and
  \begin{equation}
  A_{f}= 2 T^{(3)}_{f} - 4 Q_{f} \sin^{2}\theta_{W}, \qquad B_{f}= - 2 T^{(3)}_{f}.
  \end{equation}
  
\begin{figure}[!ht] 
\center
\includegraphics[width=80mm,height=80mm]{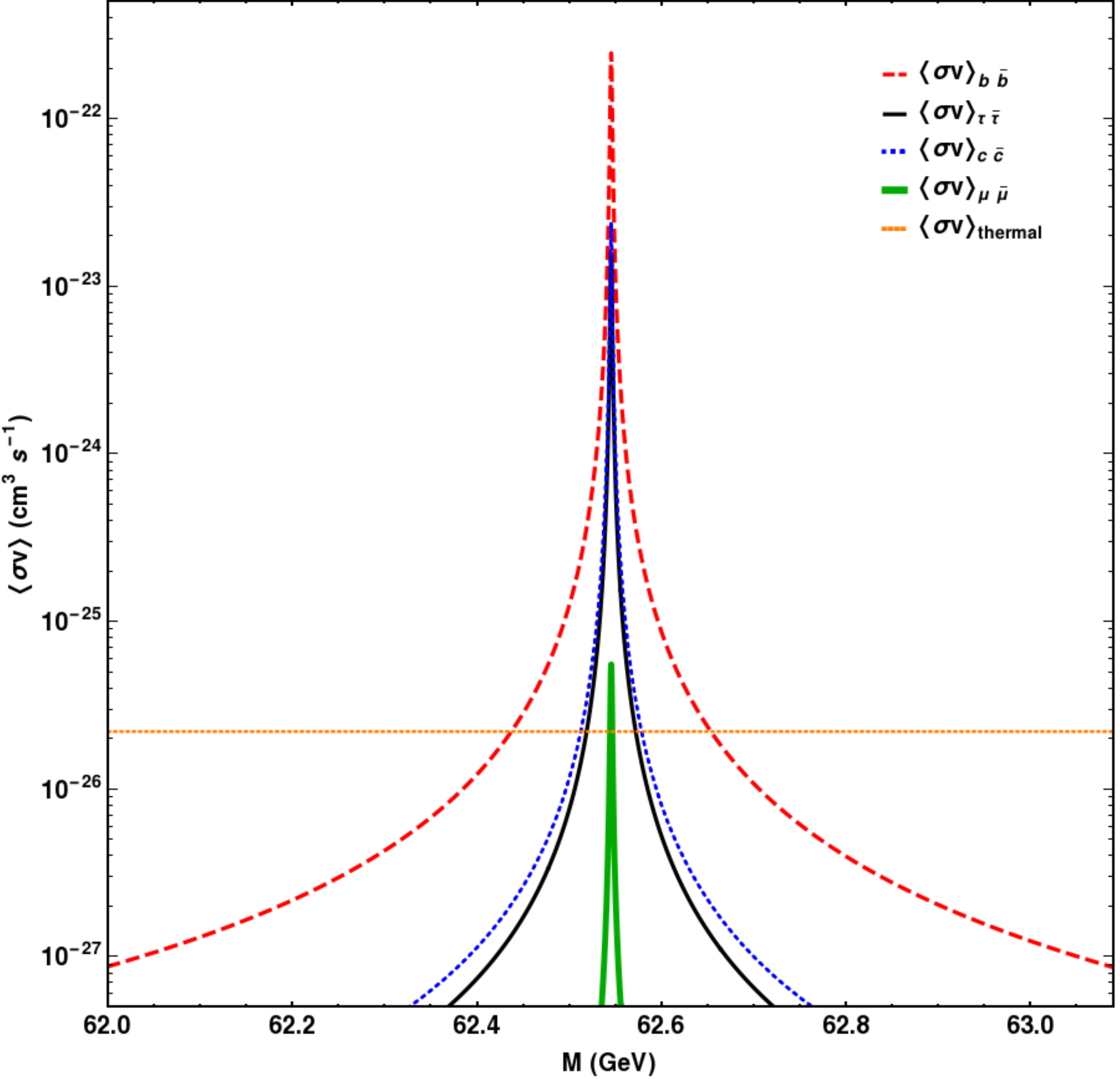}
\caption{Detail of $\langle \sigma v_{r} \rangle_{f}$ for the annihilation of SODM into fermions close to the Higgs resonance for 
$g_{s}=2\times 10^{-3}$ as a function of the SODM mass. The red line corresponds to the thermal cross 
section \cite{Steigman:2012nb}. }
\label{sigmavff}
\end{figure}

  The non-relativistic expansion of the cross section, averaged in the velocity yields

\begin{equation}
\langle\sigma v_{r} \rangle_{\bar{f}f}=\frac{N_{c}g^{2}_{s}m^{2}_{f} (M^{2}-m^{2}_{f} )^{\frac{3}{2}}} 
{12\pi M^{3}[(4M^{2}-M^{2}_{H})^{2} + M^{2}_{H} \Gamma^{2}_{H} ]}
\label{sigmavffeq}
\end{equation}
where $N_{c}=3$ for quarks and $N_{c}=1$ for leptons.  In Fig. \ref{sigmavff} we give the detail of  the velocity averaged cross sections for 
$f= \mu,\tau,c,b$, as well as the thermal cross section close to the resonance region, outside of which it takes negligible values. 
It is clear from this plot that only for SODM mass in the Higgs resonance region, 
$M\approx M_{H}/2$, we get values of the order of the thermal cross section. In the case $f=e$, even at the peak of the resonance 
we get very small values  $\langle \sigma v_{r} \rangle_{e}=1.3\times10^{-30}cm^{3}/s$ .

The sharp prediction $M\approx M_{H}/2$ for the dark matter mass suggested by the comparison of the annihilation of SODM 
into fermions with the required cross sections in the fit to the GRE data done in  \cite{Calore:2014nla}  is interesting and deserves 
a complete analysis of the photon signals produced in the annihilation of SODM into SM particles. 
The contributions of a Higgs portal to the GRE have been studied previously in the literature for dark matter with a scalar $(0,0)$, 
four-vector $(1/2,1/2)$, or spinor $(1/2,0)\oplus (0,1/2)$, space-time structures \cite{Abazajian:2014fta} \cite{Cline:2013gha} \cite{Okada:2013bna} \cite{Basak:2014sza} \cite{ Agrawal:2014oha} \cite{Duerr:2015bea} \cite{ Cuoco:2016jqt} \cite{ Sage:2016xkb}, 
mainly in the context  of models of minimal scalar dark matter to which fermions or vectors are sometimes added in several set ups to 
comply with available data on dark matter . The possibility that a dark matter with a mass 
around half the Higgs mass could explain the GRE for singlet scalar dark matter has also been pointed out in  \cite{Carena:2019pwq}
\cite{Duerr:2015bea} \cite{ Cuoco:2016jqt}\cite{ Sage:2016xkb} .  

In the next sections we analyze 
the prompt photon production from the annihilation of SODM into fermions, and the delayed photon emission from  electrons 
and muons produced in the decay of $q,\tau,c$ and $b$, which propagates in the galactic center interacting with the cosmic microwave 
background and starlight. We take into account all these contributions to find the window for the SODM mass and couplings consistent 
with the GRE and uncertainties presented in \cite{TheFermi-LAT:2017vmf}. Finally we analyze the consistency of the so obtained results 
with the constraints from direct and indirect detection for SODM in the Higgs resonance region. 

\section{Prompt photons from the annihilation of SODM into fermions} 

The couplings $g_{s},g_{p},g_{t}$ in Eq. (\ref{lag}) induce the annihilation of SODM into  final states containing photons which we 
classify as initial state radiation, internal Bremsstrahlung (or internal radiation) and final state radiation.  The simplest transitions are 
the two-body processes $\bar{D}D\to \gamma R$ 
with $R=\gamma,Z^{0}$ whose amplitudes are ${\cal O}(g^{2}_{t})$ or $R=H$ which is ${\cal O}(g_{t}g_{s},g_{t}g_{p})$. Considering 
non-perturbative QCD corrections in general $R$ in these processes can convert to quarkonium states resonances 
$\bar{Q}Q[^{2S+1}L_{J}]$ producing also $\gamma$-quarkonium final two body states. These two body processes 
yield photons with energies in a narrow energy window related to the width of the resonance $R$ and centered at 
$\omega=M(1-\frac{M^{2}_R}{4M^{2}})$. These contributions dominate the related three-body final state processes obtained 
considering the decay of $R$ into two particles, which have a continuous photon spectrum produced when $R$ is off-shell. This is the 
so-called initial state radiation. Since the three-body process in general includes the two body transitions as resonant processes we 
consider here the general case of three body transitions containing a photon in the final state. For three body transitions it is also 
possible that the SODM annihilates into a pair of particle-antiparticle with the subsequent emission of a photon, this is final state radiation. 
It is also well possible that the exchanged particle $R$ decay into another particle $R^{\prime}$ emitting a photon with the 
subsequent decay of $R^{\prime}$ into two final particles in the so called internal radiation or internal bremsstrahlung. The corresponding 
diagrams for SODM transitions $\bar{D}D\to \bar{f}f\gamma$, where $f$ stands for a fermion, are depicted in Fig. \ref{Threebody}.

\subsection{Initial state radiation}
\begin{figure}[h!]
\centering
\includegraphics[width=100mm,height=100mm]{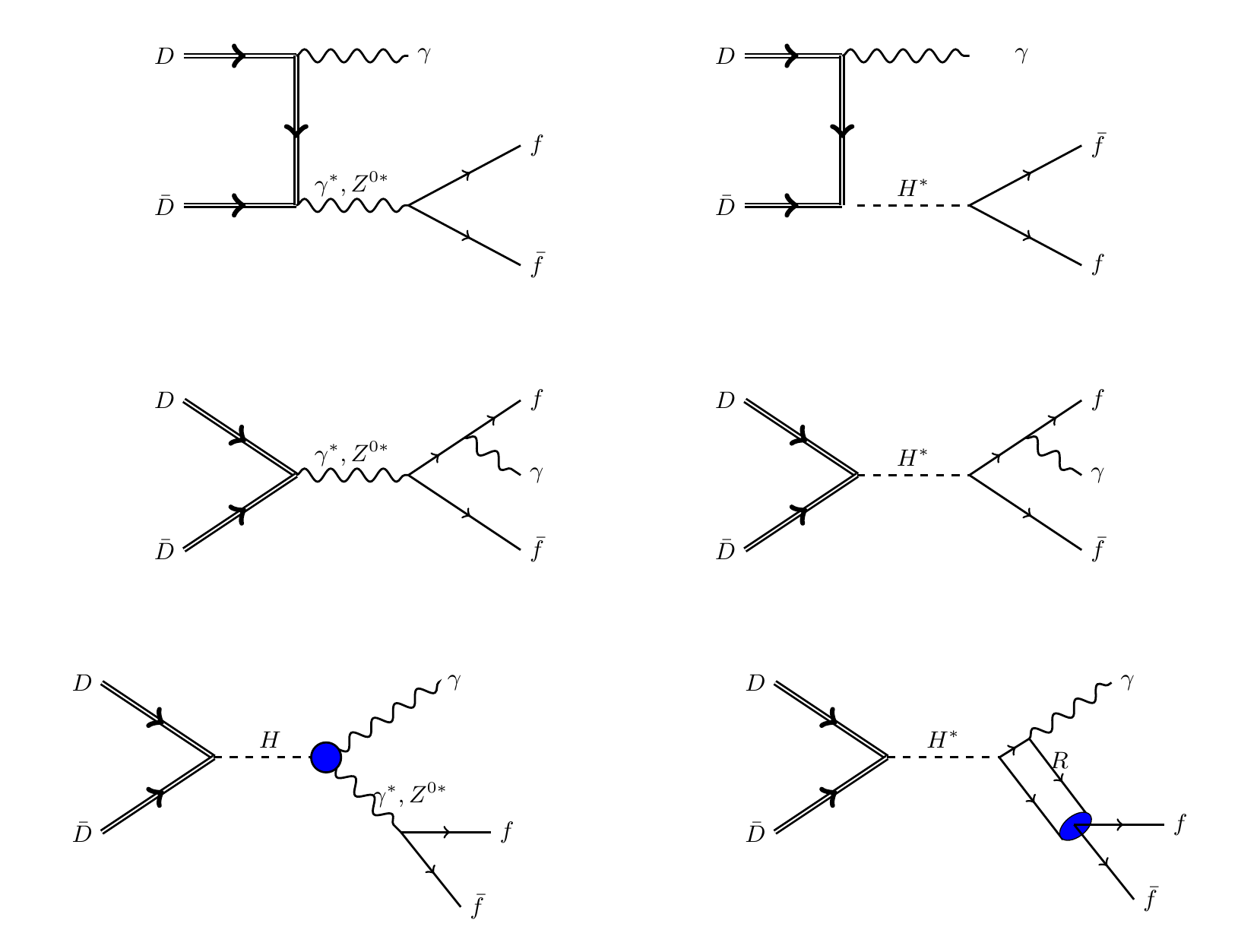}
\caption{Anihilation of SODM into three body final states containing a photon. 
There is an additional diagram for each one shown here. }
\label{Threebody}
\end{figure}

Initial state radiation is induced at tree level by the first two diagrams in Fig. \ref{Threebody}. 
It is well known that initial state radiation yield spectrums with shape similar to the GRE when there are resonant effects involved in 
the process. In this case the resonant effects translates into wider peaks in the photon spectrum (see eg. 
\cite{LucioMartinez:1994yu} for these effects in colliders at low energies).
The first diagram in Fig. \ref{Threebody}  ( $\gamma$ and $Z$ exchange) yield contributions of order ${\cal O}(v^{2}_{r})$ and 
initial state radiation is dominated by the Higgs exchange in the second diagram. We obtain  
\begin{align}
 \frac{d \langle \sigma v_{r} \rangle_{isr}}{d\omega}  
 &=\sum_{f}\frac{N_{c} \cos^2\theta_{W} g^2_{t} m^2_{f} \omega  (M-\omega )
   \left(1-\frac{m^2_{f}}{M (M-\omega)}\right)^{3/2} }{72 \pi ^3 M^5}  \nonumber \\
  & \times \frac{9 g^2_{p} \omega ^2+g^2_{s}(2 M+\omega )^2 }
  { \left(4 M (M-\omega)-M_H^2\right)^2+\Gamma _H^2 M^2_H},
\end{align}
where the sum runs over all kinematically allowed SM fermions.  As expected, for 
$M>M_{H}/2$ the photon spectrum has a 
 bump at energies corresponding to di-fermion invariant mass close to the Higgs resonance. The location of this bump depends 
 on the dark matter mass and for $M=64~GeV$ it coincides with the GRE bump at $3~GeV$.
However, this is an ${\cal{O}}(g^{2}_{t}g^{2}_{s})$ contribution to the cross-section. The coupling $g_{t}$ is severely constrained 
to $g_{t}\le 2\times  10^{-4}$ by the XENON1T results on direct detection for SODM mass of the order of $100~GeV$ 
\cite{Hernandez-Arellano:2018sen}, which makes the initial state radiation very small compared with the GRE data. 

\subsection{Final state radiation}
\begin{figure}[!ht]
\center
\includegraphics[scale=0.5]{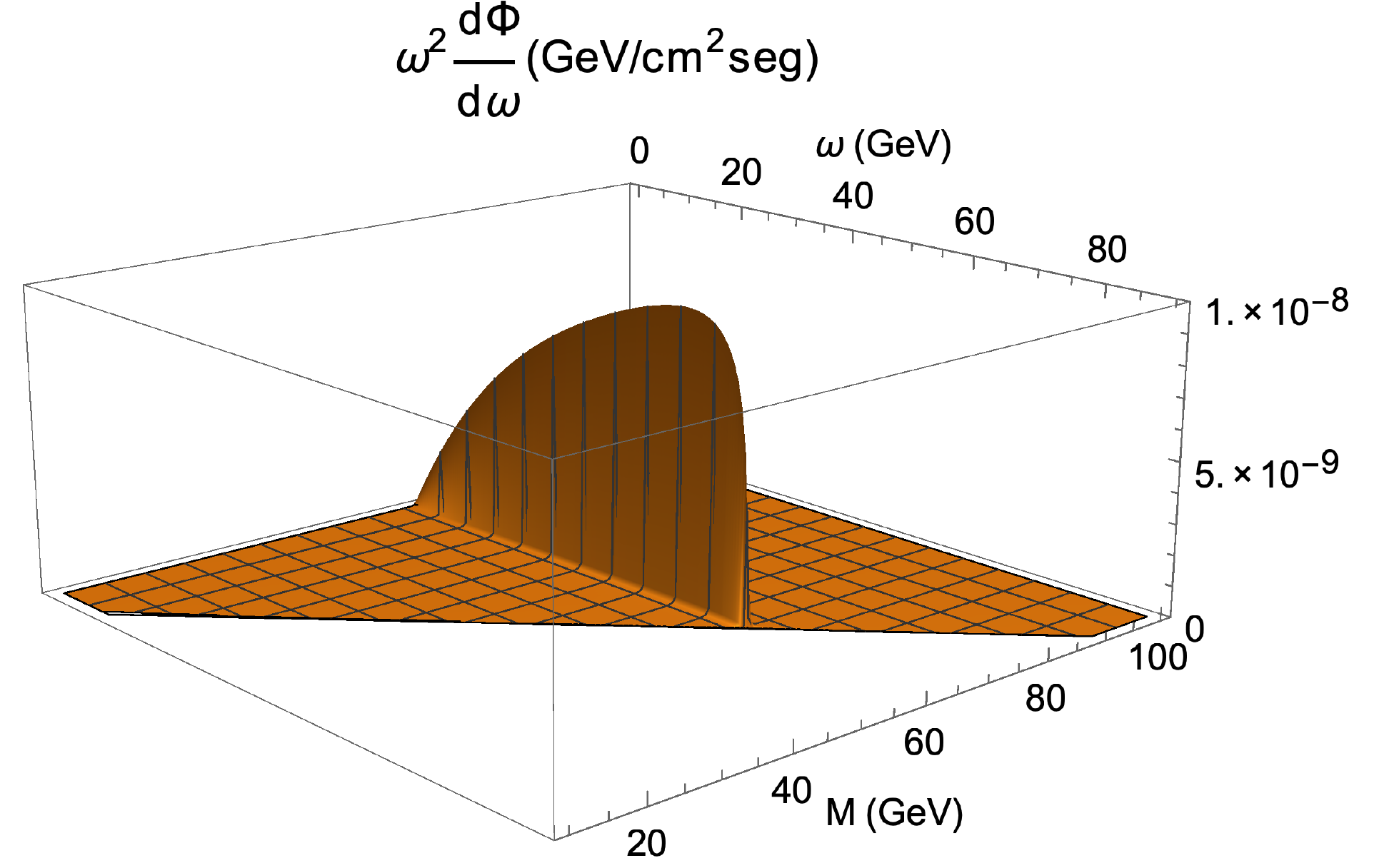}
\caption{Differential photon flux for final state radiation as a function of $\omega, M$ for $g_s = 5\times10^{-3}$.}
\label{Resfsr}
\end{figure}

Prompt photons can be also emitted by the final fermions in the the reaction $\bar{D}D\to \bar{f}f\gamma$. These contributions 
are given by the next two diagrams in Fig. \ref{Threebody}. The $\gamma$ and $Z$ exchange and the Higgs exchange 
with pseudoscalar coupling $g_{p}$  are  ${\cal O}(v^{2}_{r})$. 
The leading contributions are given by the diagrams with the Higgs exchange and scalar coupling. The differential averaged 
cross section is
\begin{align}
 \frac{d \langle \sigma v_{r} \rangle_{fsr} }{d\omega}  
 &=\sum_{f} \frac{ N_{c} \alpha Q^2_{f}  g^2_{s} m^2_{f}}{ 6\pi^2 M^4 \omega  \left(\left(4 M^2-M_H^2\right)^2+\Gamma _H^2 M_H^2\right)}  \nonumber \\
 & \left[   \left[ 2 M^3(M- \omega) +M^2 \left( \omega^2-3 m^2_{f} \right)+2 M m^2_{f} \omega +m^4_{f} \right] 
 ArcTanh \sqrt{1-\frac{m^2_{f}}{M(M- \omega )}}   \right. \nonumber \\
& \left. - M (M^2-m^{2}_{f}) (M-\omega )  \sqrt{1-\frac{m^2_{f} }{M(M-\omega )}} ~ \right].
\label{dsvdwfsr}
\end{align}
The corresponding differential photon flux as a function of the dark matter mass is shown in Fig.\ref{Resfsr} . 
These contributions are also resonant but in this case the resonant effects occur exactly at 
\begin{equation}
s=(p_{1}+p_{2})^{2}= 4M^{2}\left(1+\frac{v^{2}_{r}}{4} +{\cal O}(v^{4}_{r})\right)^{2}= M^{2}_{H}.
\end{equation}
From this plot it is clear that results for $\omega^{2}d\Phi/d\omega$ of the order of the GRE are obtained only for Higgs exchange 
in the resonance region. Since dark matter is non-relativistic, this requires $M\approx M_{H}/2$. 

The direct emission of a photon  by the fermion produced in SODM annihilation is only one of the many processes yielding prompt 
photons. For $f=\mu,\tau, q,c,b$ additional prompt photons can be produced by the decay products in the case of leptons or by the 
jet of particles produced in the hadronization of quarks. We expect these effects to modify substantially our results in 
Eq. (\ref{dsvdwfsr}) for all fermions except  for $f=e,\mu$ which do not have hadronic decays and receive only modifications from 
suppressed higher order electroweak radiative corrections.

\begin{figure}[ht]
\center
\includegraphics[width=140mm,height=90mm]{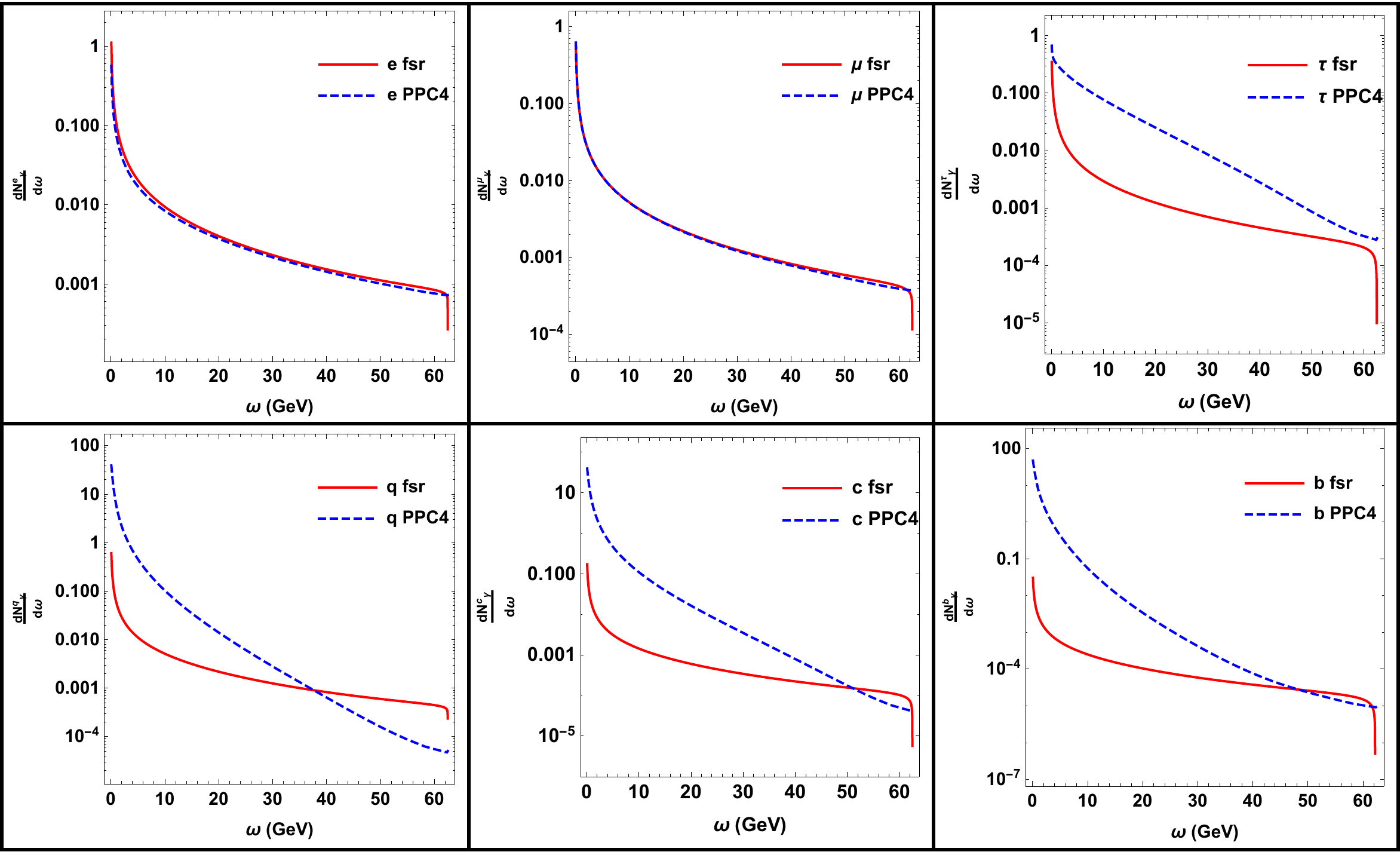}
\caption{Comparison of the tree level result in Eq. (\ref{dsvdwfsr}) for $\frac{dN^{f}_{\gamma}}{d\omega}$ versus the results using PPC4DMID 
tabulated spectrum, for leptons and quarks, as a function of $\omega$ for $M=62.5~GeV$.}
\label{specfsrvsppc4}
\end{figure}

We calculate the complete prompt photon flux for $f=e,\mu,\tau,q,c,b$ using the tabulated spectrum defined as
provided by DARKSUSY \cite{Gondolo:2004sc} and PPC4DMID \cite{Cirelli:2010xx}, including radiative corrections \cite{Ciafaloni:2011sa}. 
We check the consistency of results using both packages and use the spectrum given by the direct photon emission 
by electrons and muons in Eq.(\ref{dsvdwfsr}) to cross-check results. The comparison of the direct photon emission in Eq. (\ref{dsvdwfsr}) 
versus the tabulated spectrum from PPC4DMID are shown in Fig. \ref{specfsrvsppc4}, which confirms our expectations.

\begin{figure}[ht]
\center
\includegraphics[scale=0.4]{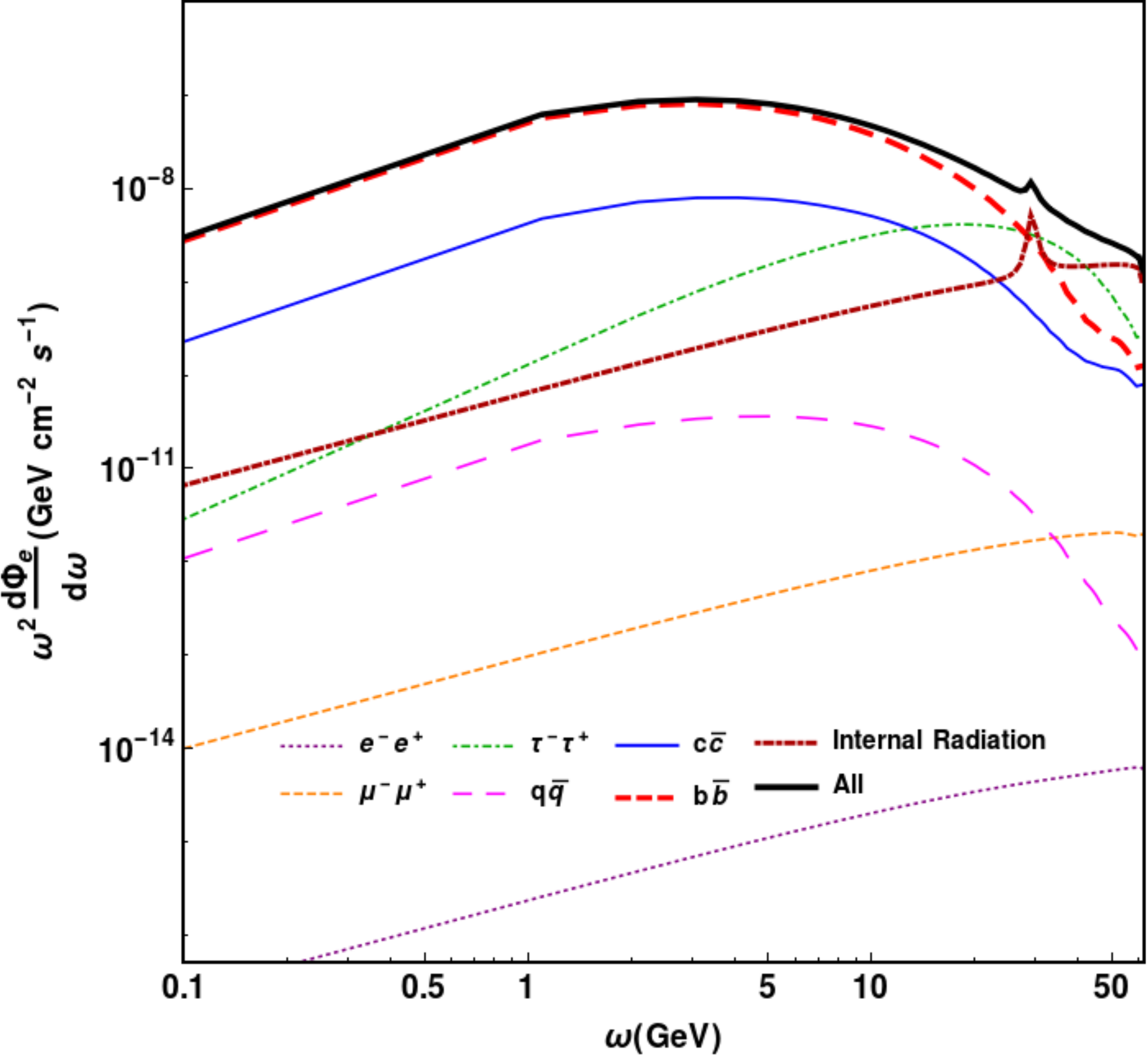}
\caption{Differential flux for prompt photons from the annihilation of SODM into fermions for $M=62.49~GeV$ and $g_{s}=10^{-3}$. We use 
$M_{H}=125.09$ in the computation of these contributions.}
\label{prompt}
\end{figure}

The prompt photon flux is calculated using the tabulated spectrum in PPC4DMID (DARKSUSY yields similar results) and the 
SODM result for  $\langle \sigma v_{r} \rangle_{\bar{f}f} $ in Eq.(\ref{sigmavffeq}) for each fermion. Our results for $g_{s}= 10^{-3}$ 
and $M=62.49~GeV$ are shown in Fig. \ref{prompt}.  In the computation 
of the flux we use the gNFW profile \cite{Zhao:1995cp}\cite{Kravtsov:1997dp} with $\gamma=1.25 $  which for the region of interest 
$\left\vert l\right\vert <10^{\circ}$ and $2^{\circ}\le \left\vert b\right\vert <10^{\circ}$ yields $J_{0}=7.12\times 10^{5}GeV^{4}/cm^{2}~seg$.

The most important contributions comes from the $b$ channel followed by the $c$ and $\tau$ which 
become competitive at high photon energies. Prompt photon flux from electrons muons and light quarks turn out to be negligible.

\subsection{Internal Bremsstrahlung}
The internal state radiation is given by the last two diagrams of Fig. \ref{Threebody}. The first of these diagrams involve the 
$H\to \gamma \gamma$ and $H\to Z\gamma$ transitions which takes place at one-loop level in the standard model. 
The $H\gamma\gamma$ and $H\gamma Z$ three-point functions have been calculated in the 
literature \cite{Bergstrom:1985hp}\cite{Barroso:1985et} \cite{Bonciani:2015eua}. Considering only the parts contributing 
to our processes, the SM yields the following effective interactions
\begin{equation}
{\cal L}_{eff}=H[G_{\gamma\gamma}  F^{\mu\nu}F_{\mu\nu} +G_{Z\gamma} F^{\mu\nu}  Z_{\mu\nu}  ],
 \end{equation}
where $G_{\gamma\gamma}, G_{Z\gamma}$ are the corresponding form factors. We will follow a phenomenological approach here, 
and normalizing the form factors as $G_{\gamma\gamma}= \frac{g_{\gamma\gamma}}{M_{H}}$, 
$G_{Z\gamma}= \frac{g_{Z\gamma}}{M_{H}} $, will extract the couplings from the measured branching rations 
$BR[H\to \gamma\gamma]=2.27\times 10^{-3}$, $BR[H\to Z\gamma]=1.53\times 10^{-3}$ \cite{Tanabashi:2018oca} to obtain 
 $g_{\gamma\gamma}=1.91\times 10^{-3}$, $g_{Z\gamma}=3.30\times 10^{-3}$. These couplings correspond to the form 
 factors for on-shell momentum. As it will be shown below, only the resonant processes produce sizable contributions for most of the 
 channels relevant in this work, hence this approximation is justified for our purposes.

For the sequential decay with $\gamma$ and $Z^{0}$ 
intermediate states we obtain
  \begin{align}\nonumber
 & \frac{\langle d\sigma v_{r} \rangle_{Z^{*}}}{d\omega}= \sum_{f}
  \frac{N_{c} g^2_{Z\gamma} g^2_{s} \omega ^3 M^2_Z  \sqrt{1-\frac{m^2_{f}}{M (M-\omega )}} }
  {36 \pi ^3 M_H^2 \left(\left(4 M^2-M_H^2\right){}^2+\Gamma _H^2 M_H^2\right)}  \\
 &\times  \frac{2(A^2_{f} +B^2_{f}) M (M-\omega )+(A^2_{f}-2B^2_{f})m^2_{f}}
  {  \left(\left(4 M (M-\omega )-M_Z^2\right)^2+M_Z^2 \Gamma _Z^2\right)} , 
  \end{align}
  \begin{align}\nonumber
& \frac{\langle d\sigma v_{r} \rangle_{\gamma^{*}}}{d\omega}=  \sum_{f}
 \frac{N_{c} \alpha  g^2_{\gamma\gamma} g^2_{s} v^2 }{36 \pi ^2 M_H^2 \left(\left(4M^2-M_H^2\right)^2+\Gamma _H^2 M_H^2\right)}  \\
  &\times  \frac{\omega ^3 \left(2 M  (M-\omega )+m^2_{f}\right)\sqrt{1-\frac{m^2_{f}}{M(M -\omega) }}}{M^2(M-\omega )^2 } . 
  \end{align}
 There is also an enhancement at the Higgs resonance in these processes and 
  a double-resonant effect in the case of the $Z^{*}$ intermediate state. The last diagram in Fig. \ref{Threebody} involves 
  non-perturbative QCD effects. We calculated these contributions using the Non-Relativistic QCD effective field theory finding 
  them negligible even at the Higgs resonance. There are also contributions with the sequential decays 
  $\bar{D}D\to \gamma*,Z^{0*} \to \gamma H\to \gamma \bar{f}f$ not shown in Fig. \ref{Threebody} which are not resonant 
  and are also very small.
  
  Our results for the internal radiation are shown in Fig. \ref{prompt}. Sizable contributions to the differential photon flux 
  from $H\to \gamma Z^{*}$ transition are produced mainly at the $Z^{0}$ resonance i.e. for photon energies around 
  $\omega=\frac{M_{H}}{2}(1-\frac{M^{2}_{H}}{M^{2}_{Z}})\approx 30 ~ GeV$. The $H\to\gamma \gamma^{*}$ intermediate state 
  contributes only at the upper end of the spectrum. 
  
\section{Delayed emission: Inverse Compton Scattering contributions}
There are at least three different contributions from the delayed photon emission by ICS of propagating fermions produced in the 
annihilation of SODM: i) The propagation of electrons
produced in $\bar{D}D\to e^{+}e^{-}$, which is negligible due to the small coupling with the exchanged Higgs, $g_{Hee}=m_{e}/v$, 
or the small coupling of $\gamma$ and $Z$ to SODM (proportional to $g_{t}$) if we consider the spin portal; ii) The propagation of 
muons produced in $\bar{D}D\to \mu^{+}\mu^{-}$ which was shown in \cite{Calore:2014nla} to yield sizable contributions; iii) 
The propagation of electrons coming from the decay of leptons or hadronization of quarks, produced in $\bar{D}D\to \bar{f}f$ with 
$f=\mu,\tau,q,c,b$. 

\begin{figure}[ht]
\center
\includegraphics[scale=0.4]{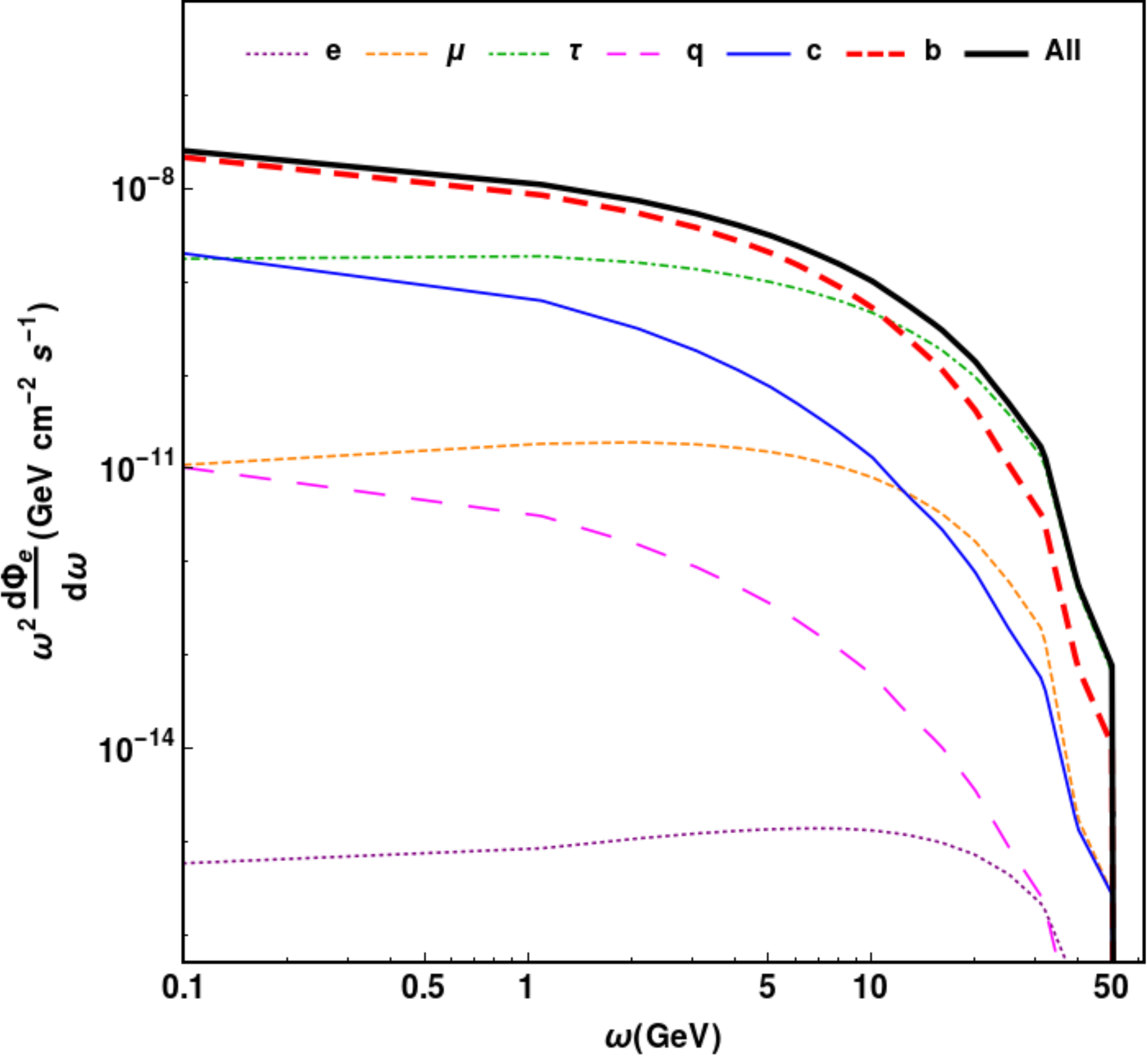}
\caption{Differential flux of delayed photons produced in the ICS of electrons off CMB and starlight, for secondary electrons produced 
in the decay of heavier leptons or hadronization of quarks  coming from the annihilation of SODM, for $M=62.49~GeV$ and 
$g_{s}=10^{-3}$. We use $M_{H}=125.09$ in the computation of these contributions.}
\label{icse}
\end{figure}

We compute the ICS of electrons by the cosmic microwave background or starlight for electrons produced in the decay of heavier 
leptons or in the hadronization of quarks using the tabulated electron spectrum and propagating models in  PPC4DMID 
\cite{Cirelli:2010xx} \cite{Buch:2015iya}.  Our results for the photon flux from the different fermions are given in Fig \ref{icse}. We warn that 
these contributions are calculated with the NFW\cite{Navarro:1995iw} density profile since the PPC4DMID tabulated spectrum is 
designed only for specific profiles not including the gNFW. However, these contributions are not very sensitive to the density profile 
and results are quite similar using the Moore \cite{Diemand:2004wh} or Einasto B \cite{Graham:2005xx} \cite{Cirelli:2010xx} profiles 
included in the PPC4DMID setup and are more alike to the gNFW used in the computation of prompt photons.  

In this case also the $b$ channel is dominant, with subdominant contributions of the $c$ and $\tau$ channels, light quarks, with 
electrons and muons yielding negligible contributions. We remark that for both prompt and delayed 
 photons, these contributions rise as we go deeper inside the resonance region, results being highly sensible to the specific value 
of the SODM mass.  

There are also contributions to the delayed  photon flux from the ICS of muons produced in $\bar{D}D\to \mu^{+}\mu^{-}$ calculated in \cite{Calore:2014nla}. 
 However, for the SODM mass window required by the GRE data when we include all the contributions, the cross section  
$\langle \sigma v_{r} \rangle _{\mu}$ is about three orders of magnitude below the thermal cross section needed for this mechanism to 
yield a sizable contribution and the corresponding flux turns out to be very small.

  \section{Final results}
  
\begin{figure}[!ht] 
\center
\includegraphics[scale=0.4]{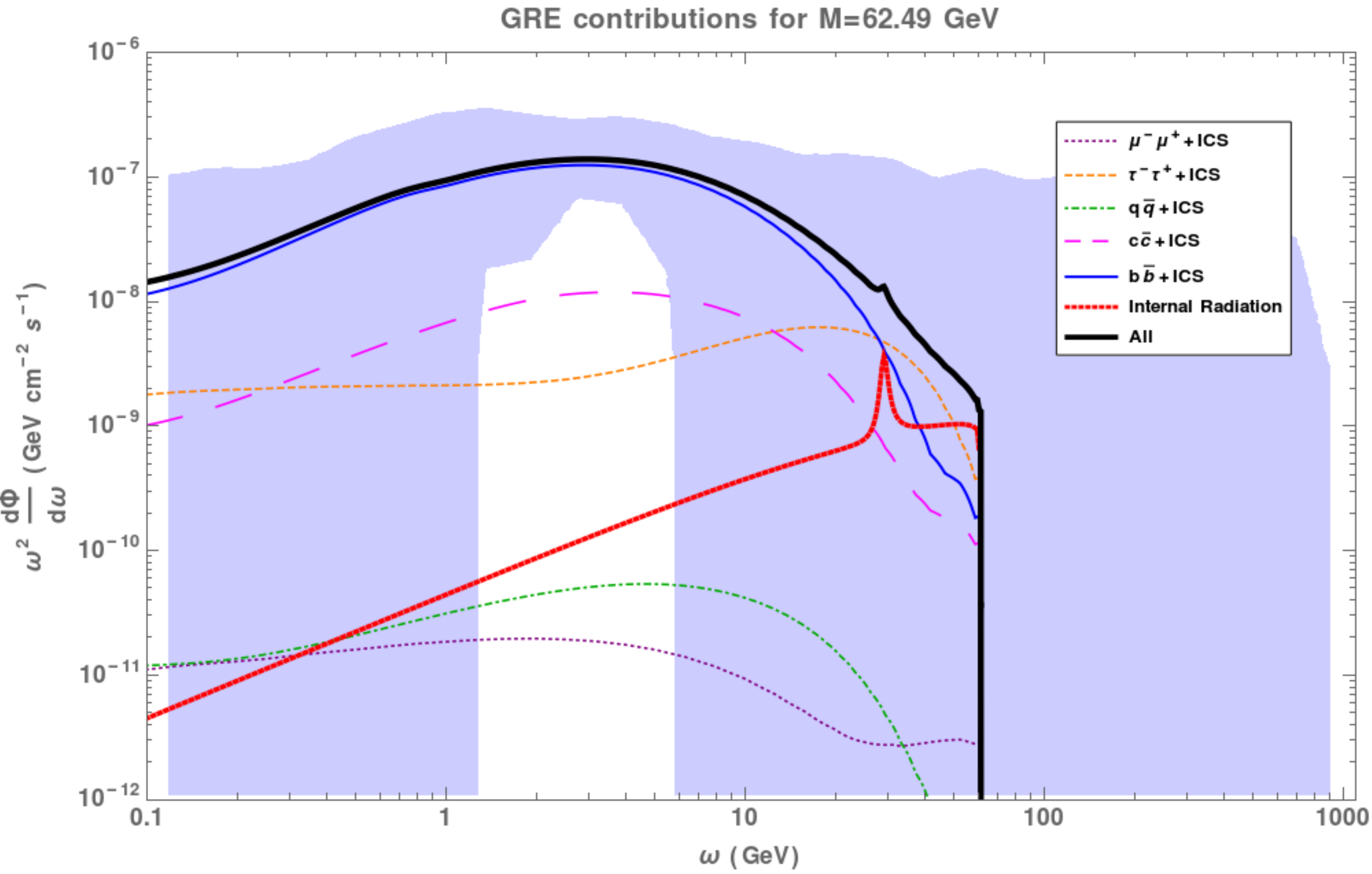}
\caption{Differential flux as a function of $\omega$ including all the contributions discussed in this paper, for $M = 62.49 \text{ GeV}$ 
and $g_s = 9.81 \times 10^{-4}$.}
\label{fluxall}
\end{figure}

 Including all the contributions described in the previous section we obtain the results shown in Fig. \ref{fluxall}. 
 Although these results in principle depend on the SODM mass $M$ and on the couplings of the model $g_{s}, g_{p},g_{t}$,  
 the spin portal ($g_{t}$) and pseudoscalar ($g_{p}$) couplings yield negligible contributions to the photon flux which depends only on 
 the parity-conserving Higgs-SODM interaction coupling $g_{s}$ and $M$. 
 
 As to the size of the individual contributions, we have a similar amount of prompt and delayed photons, but prompt photons are dominant 
 for $\omega > 0.3 ~GeV$. In both cases, the main contributions come from the $\bar{D}D \to \bar{b}b$ 
 annihilation with small but sizable contributions from the $\tau^{+}\tau^{-}$ and $\bar{c}c$ channels. Among prompt photons most of the flux
 comes from final state radiation. Internal radiation yields competitive contributions only at the upper end of the spectrum. 

In Fig.\ref{fluxall}, we show also the shadowed band allowed by the uncertainties on the GRE data obtained 
in \cite{TheFermi-LAT:2017vmf}. Considering these uncertainties we obtain the values 
$g_{s}\in [0.98, 1.01] \times 10^{-3}$ and $M\in [62.470,62.505]~GeV$ consistent with the GRE data. These sharp set of values yield 
definite predictions for other observables of dark matter and we must ensure that existing constraints are satisfied. In the next section we 
work out the corresponding results for constraints from relic density, direct and indirect searches for dark matter.

\section{Constraints from direct and indirect detection experiments at the Higgs resonance}
A global analysis of the predictions of the 
formalism for results on the relic density, direct searches (XENON1T upper bound for the dark matter-proton cross section $\sigma_{p}$) 
and indirect searches (upper bounds on the annihilation of dark matter into $\bar{b}b, \tau^{+}\tau^{-}, \mu^{+}\mu^{-}$)  was 
performed in \cite{Hernandez-Arellano:2018sen}, finding consistency with available data. However, the sharp result on the SODM 
mass imposed by the GRE, requires a deeper analysis for a SODM mass at the Higgs resonance because, on the one hand, the 
non-relativistic approach used in the calculation of the dark matter relic density is well known to break down in the presence of 
resonances, and on the other hand, the cross-sections for the annihilation of dark matter 
dramatically grow at the Higgs resonance and we must ensure that the annihilation cross section to fermions, specially 
to $\bar{b}b$, is consistent with present upper limits. 
 
\subsection{Direct detection}
The XENON1T upper bound for $\sigma_{p}$ at $M= M_{H}/2$ is $ \sigma_{p}\leq 9.86\times 10^{-47} cm^{2} $.     
This upper bound requires $g_{s}\leq 5.12\times 10^{-3}$ for the Higgs portal or $g_{t} \leq 1.7\times 10^{-5}$ for the spin portal. 
The window $g_{s}\in [0.98, 1.01] \times 10^{-3}$  imposed by the GRE data is consistent with the upper bound on $g_{s}$ imposed 
by XENON1T. 
\subsection{Relic density} 

Our results for the dark matter relic density in \cite{Hernandez-Arellano:2018sen} must be refined if $M\approx M_{H}/2$, since 
we are at the Higgs resonance and it has been shown previously in a general analysis that the naive calculation of the 
dark matter relic density using the non-relativistic expansion can fail in the presence of resonances \cite{Griest:1990kh}.
The scope of this failure depends on the specific values of the resonance mass and width, but it can be important well 
beyond the resonance region. Our model allows to test how important it can be for the Higgs resonance. 

The calculation of the relic density requires to consider the full thermal average cross-section $\langle \sigma v_{r} \rangle (x)$ 
for the annihilation of dark matter into standard model states. For masses $M\approx M_{H}/2$, dark matter annihilates only into 
$\bar{f}f$, $\gamma\gamma$ and $Z^{0}\gamma$. The cross section for the annihilation into light fermions 
is given in Eq.(\ref{sigmavffcomp}). 

\begin{figure}[h!]
\centering
\includegraphics[scale=1]{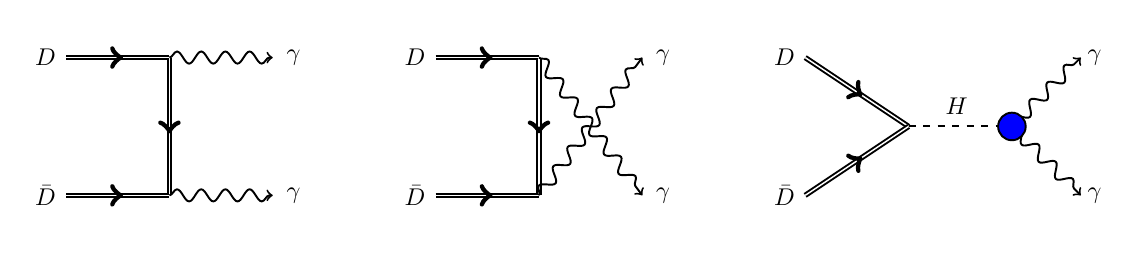}
\caption{Anihilation of SODM into two photons. The last diagram represents the one loop contributions in the SM.}
\label{Twophotons}
\end{figure}

The leading contributions for SODM annihilation into  $\gamma\gamma$ and $Z^{0}\gamma$ via Higgs exchange are 
shown in Fig. \ref{Twophotons} and involve next to leading order SM contributions inducing the $H\to \gamma\gamma$ and  
$H\to \gamma Z$ transitions. The spin portal contributions generated by the $g_{t}$ coupling turn out to be very small and the 
pseudoscalar Higgs portal induced by $g_{p}$ are ${\cal O}(v^{2}_{r})$, thus  we keep only the scalar Higgs portal contributions 
given by $g_{s}$.

\begin{figure}[!ht] 
\center
\includegraphics[scale=0.3]{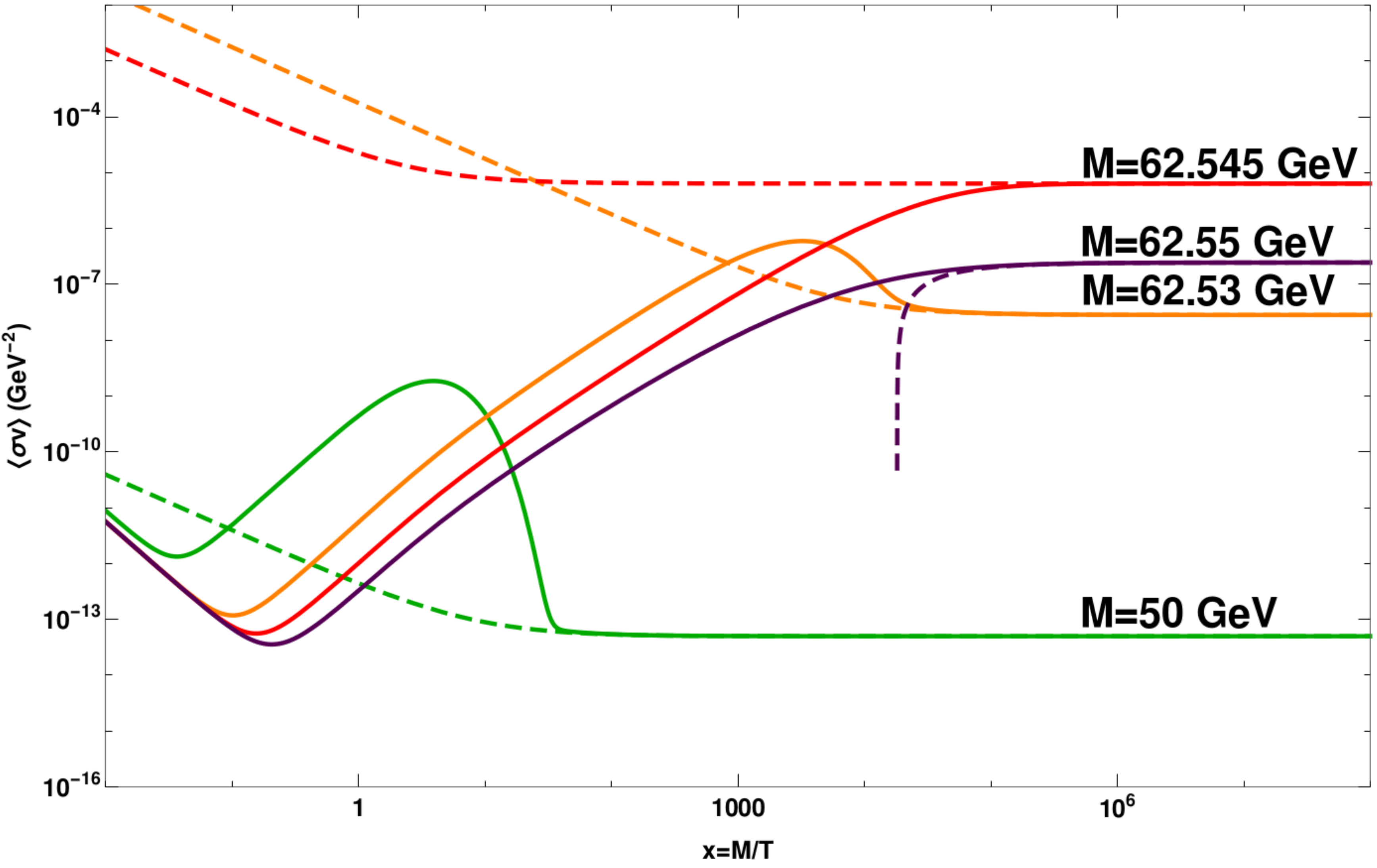}
\caption{Thermal average cross-section (solid) and comparison with the non-relativistic expansion (dashed), for different values of the SODM 
mass. }
\label{sigmav}
\end{figure}

\begin{figure}[!ht] 
\center
\includegraphics[scale=0.4]{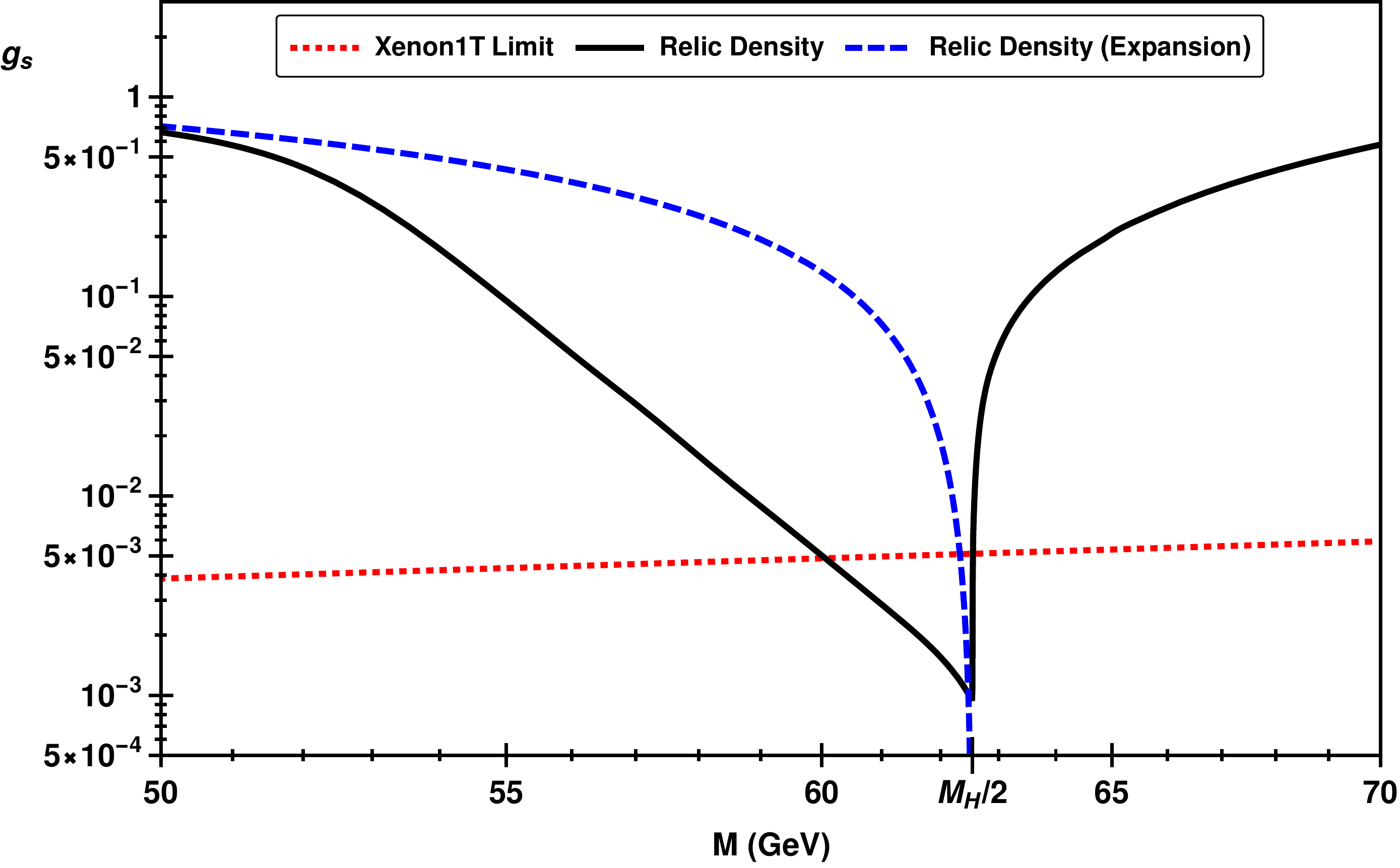}
\caption{Values of the coupling $g_s$ and SODM mass $M$ consistent with the measured relic density near the Higgs resonance 
obtained without the $v^{2}_{r}$ expansion (continuous line). The dashed line correspond to the conventional non-relativistic expansion. The short-dashed line is the upper bound on $g_{s}$ imposed by XENON1T results on $\sigma_{p}$\cite{Hernandez-Arellano:2018sen}. }
\label{gsres}
\end{figure}

A straightforward calculation of $\bar{D}D\to \gamma\gamma, ~ Z^{0}\gamma$ 
induced by Higgs exchange yields
\begin{align}\label{gg}
( \sigma v_r )_{\gamma \gamma} &= \frac{g_{\gamma \gamma}^2 g_s^2 v^2 s^2  (6 M^4 - 4M^2 s + s^2)}{288\pi M^4 M_H^2  
(s-2M^2)[ (s-M_H^2)^2 + M_H^2 \Gamma_H^2 ] } , \\
( \sigma v_r )_{Z \gamma} &= \frac{g_{Z\gamma }^2 g_s^2 v^2 (s-M_Z^2)^3 (6 M^4 - 4M^2 s + s^2)}{144 \pi  M^4 M_H^2 
(s-2M^2) s [ (s-M_H^2)^2 + M_H^2 \Gamma_H^2 ]} . 
\end{align}
We use these results to calculate numerically the complete 
thermal average cross section. Our results compared with those obtained with the 
non-relativistic expansion are shown in Fig. \ref{sigmav}, where it can be seen that even for values of $M$ far from the resonance there are 
important differences and, at the resonance, these differences are dramatic and extend to the highly non relativistic regime.
We use the complete function $\langle \sigma v_{r} \rangle (x)$ to solve the freezing condition and the Boltzmann equation for the relic 
density following the conventional procedure used in \cite{Hernandez-Arellano:2018sen}. 
The freezing temperature is still around $x_{f}\approx 25$. The corresponding values of the coupling $g_{s}$ and the dark matter mass $M$ 
consistent with the measured relic density are shown in Fig. \ref{gsres}. Consistency with the XENON1T upper bounds on $\sigma_{p}$ 
holds only for a mass $M\in [60.05, 62.96]~ GeV$ and restricts the scalar coupling to values in the $g_{s}\in[0.95, 5.15]\times 10^{-3}$ window. 
These values are consistent with the GRE data which requires $g_{s}\in [0.98, 1.01] \times 10^{-3}$ and $M\in [62.47,62.50]~GeV$, but the measured relic density correlates the values of these parameters to the solid curve in Fig. \ref{gsres}. 

\subsection{Annihilation into $\mu^{+}\mu^{-}$, $\tau^{+}\tau^{-}$ and $\bar{b} b$}

Upper bounds for the $\mu$ channel were obtained in \cite{Bergstrom:2013jra} as a function of the dark matter mass using the 
AMS02 data on the positron fraction in primary cosmic rays \cite{Aguilar:2013qda}. For $M=62.5~ GeV$ the upper limit is 
$\langle \sigma v_{r} \rangle_{\mu} \le 8.96 \times 10^{-26} cm^{3}/seg$ and this value is stable across
the resonance region. For SODM the largest values of this cross section compatible with the GRE data is obtained for  
$M= 62.505 ~GeV$ for which the measured relic density requires $g_{s} = 9.81\times10^{-4}$.  A calculation 
of the SODM prediction with these values yields  $\langle \sigma v_{r} \rangle_{\mu} \le 8.30 \times 10^{-30} cm^{3}/seg$ well 
below the experimental upper bound.

As for the $\tau$ channel the upper limit obtained in  \cite{Drlica-Wagner:2015xua} for $M=62.5~ GeV$ is 
$\langle \sigma v_{r} \rangle_{\tau} \le 1.2 \times 10^{-26} cm^{3}/seg$, while a calculation in our formalism of the largest 
value compatible with the GRE data yields $\langle \sigma v_{r} \rangle_{\tau} = 2.42\times10^{-27}~ cm^3/s$, consistent with the
experimental upper bound.

\begin{figure}[!ht] 
\center
\includegraphics[scale=0.3]{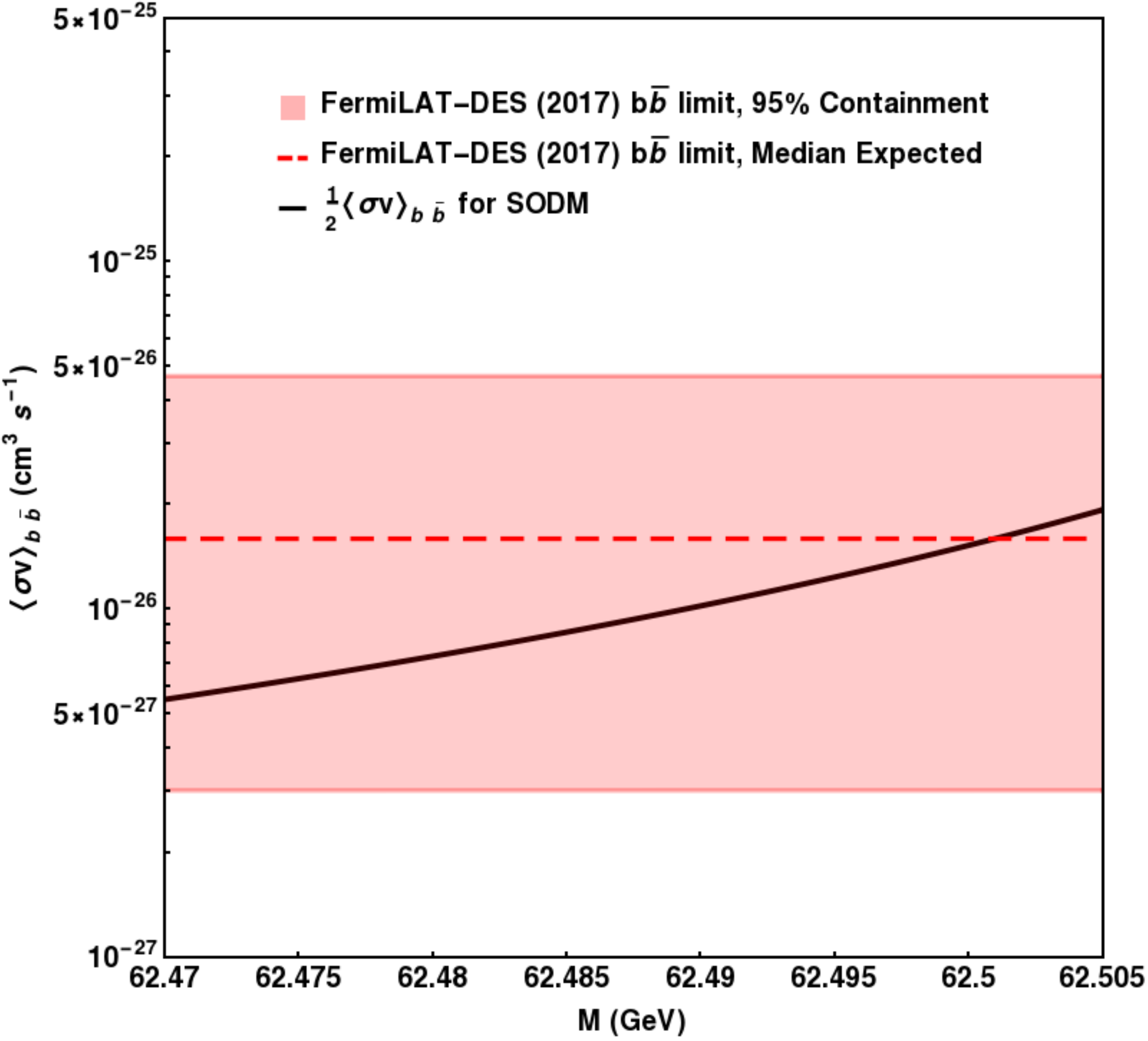}
\caption{Cross sections for the annihilation of SODM into $ \bar{b} b$ in the mass window compatible with the GRE data (solid line). 
The dashed line correspond to the Median Expected and the shadowed band to the $95\% $ containment region obtained in 
\cite{Drlica-Wagner:2015xua}. }
\label{blimit}
\end{figure}

The $b$ channel deserves a closer analysis since our results are closer to the upper bounds. In Fig. \ref{blimit} we plot the 
median expected limit and the region for $95\%$ containment  obtained in  \cite{Drlica-Wagner:2015xua} and our results 
for $\frac{1}{2}\langle \sigma v_{r} \rangle_{\tau}$ (the $\frac{1}{2}$ factor is required since the upper bounds are obtained for 
self-conjugated dark matter) . Although our results are 
consistent with the $95\%$ containment band for the whole mass window compatible with the GRE, and below the median 
expected limit for a narrower mas window $M\in [62.470, 62. 480]~GeV$, it is important to analyze this compatibility in detail since, 
on the light of the results in the present work, this channel is a very promising place to look for signals of SODM.     
In this concern, it is important to remark that the Higgs is a very narrow resonance ($\Gamma_{H}/M_{H} =3.2\times 10^{-5}$), 
its energy resolution is not easy and even in collider experiments it is still a pending task.

The upper bounds as a function of the dark matter mass are obtained in  \cite{Drlica-Wagner:2015xua}  \cite{ Fermi-LAT:2016uux} 
from a combined analysis of the data on the energy flux from 45 dwarf spheroidal galaxies (dSph). Only 19 of these 45 targets
have a $J$-factor derived from experimental data on stellar dynamics, a work done in  \cite{Geringer-Sameth:2014yza}, the fluxes for the
remaining 26 targets are estimated from the empirical relation between the flux and the inverse square of the distance satisfied by 
the dSphs with flux derived from data on stellar dynamics. 

For each target, the likelihood analysis performed for the photon energy flux consider 24 logarithmically spaced energy bins in the 
energy region from $500 ~MeV$ to $500 ~GeV$ using the likelihood function data provided in the supplementary material of 
\cite{ Fermi-LAT:2016uux}. Then  the bin-by-bin upper bound for the photon flux excess at the $95\%$ confidence level is calculated 
following the likelihood formalism described in \cite{Ackermann:2013yva}. These results are then used to estimate the upper bound on 
$\langle \sigma v_{r} \rangle_{b}$ for a given mass $M$ using the standard model results for the photon spectrum from $b$ quarks 
for each target and a combined likelihood analysis of these results yields the plot in Fig. 9 of \cite{ Fermi-LAT:2016uux}.     

\begin{figure}[!ht] 
\center
\includegraphics[scale=0.13]{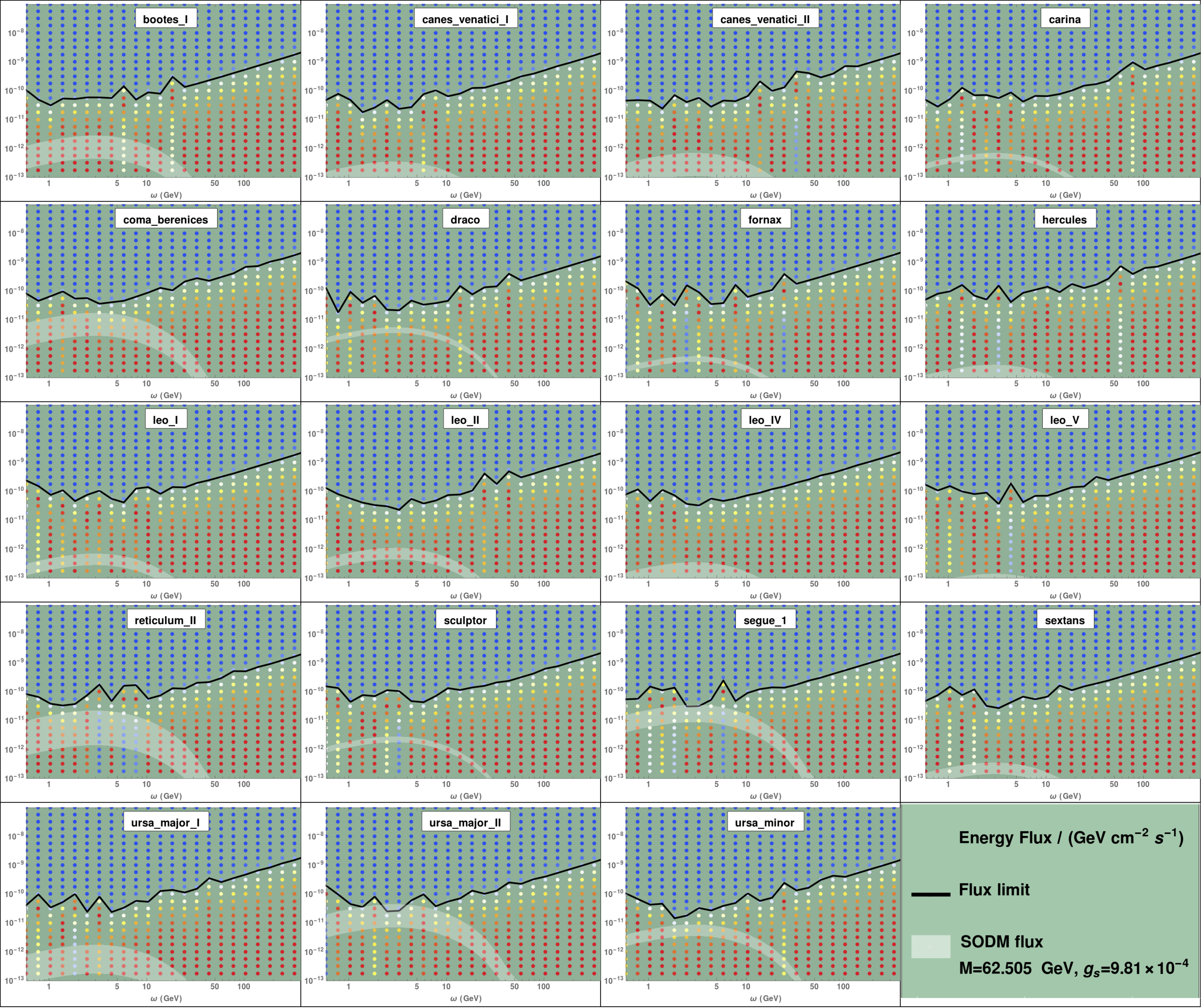}
\caption{Energy flux (in units of $GeV~cm^{-2}~s^{-1}$) from each of the 19 targets  whose J-factor was derived from stellar kinematics in \cite{Geringer-Sameth:2014yza} and 
were used in \cite{ Fermi-LAT:2016uux} to extract the upper bounds for the annihilation of dark matter into $\bar{b}b$.  
The colored dots represent the values of the likelihood function, where red is a higher value and blue is lower. The black solid line 
is the energy flux upper limit at 95 \% confidence level obtained from the bin-by-bin likelihood functions. The shaded white area 
represents the energy flux for SODM annihilating into b-quark pairs, using the measured J-factor and its uncertainty, for 
$M=62.505$ GeV and $g_s = 9.81\times 10^{-4}$.}
\label{Likelihood}
\end{figure}

 In order to make a direct comparison of our results with experimental data, for each target, we reproduce the bin-by-bin upper 
 bound for the photon flux excess at the $95\%$ confidence level using the likelihood function data provided in the supplementary 
 material of \cite{ Fermi-LAT:2016uux} and  the likelihood formalism in \cite{Ackermann:2013yva}.  Then we calculate 
the predictions of the formalism for the photon energy flux for each of the 19 targets. For given values of $M$ and $g_{s}$, we obtain 
the energy flux integrating the $\bar{D}D\to\bar{b}{b}$ contribution to $\omega\frac{d\Phi}{d\omega}$  
bin-by-bin, where we use the J-factor of the considered target. We extrapolate these points to obtain the photon energy flux as a 
function of the photon energy. In Fig. \ref{Likelihood} we show our results for the largest contribution obtained in our formalism 
consistent with the GRE data, corresponding to $M=62.505$ and $g_{s}=9.81\times 10^{-4}$. We show also in Fig. \ref{Likelihood} 
the band corresponding to the uncertainties in the measured J-factor of the target.  In this plot we display only results for the 19 targets 
with $J$ factors derived from stelar dynamics data, but similar results are obtained for the rest of the targets.
The photon energy flux from SODM annihilation into b-quark pairs, for each target, turns out to be smaller than the limits obtained 
using the  bin-by-bin likelihood functions provided by FermiLAT-DES in the
supplementary material of Ref. \cite{ Fermi-LAT:2016uux}. 

\subsection{Dark matter annihilation into two photons.}

There are stringent upper bounds on  $\langle \sigma v_{r} \rangle_{\gamma\gamma}$ by FermiLAT \cite{Ackermann:2015lka} 
and HESS  \cite{Abdallah:2018qtu} collaborationsm, which for $M\approx M_{H}/2$ are 
$\langle \sigma v_{r} \rangle_{\gamma\gamma}\le 6.75 \times 10^{-29} cm^{3}/seg$. 

The  averaged cross section for the annihilation of SODM  into two photons is given in Eq.(\ref{gg}).  A straightforward calculation 
to leading order in $v^{2}_{r}$ yields 
\begin{equation}
  \langle \sigma v_{r} \rangle_{\gamma\gamma} = 
  \frac{g^2_{\gamma\gamma} g^2_{s} M^2 v^2}{6 \pi  M_H^2 \left(\left(4 M^2-M_H^2 \right)^{2} + M_H^2 \Gamma _H^2 \right)} .
\end{equation}

\begin{figure}[h!]
\centering
\includegraphics[scale=0.3]{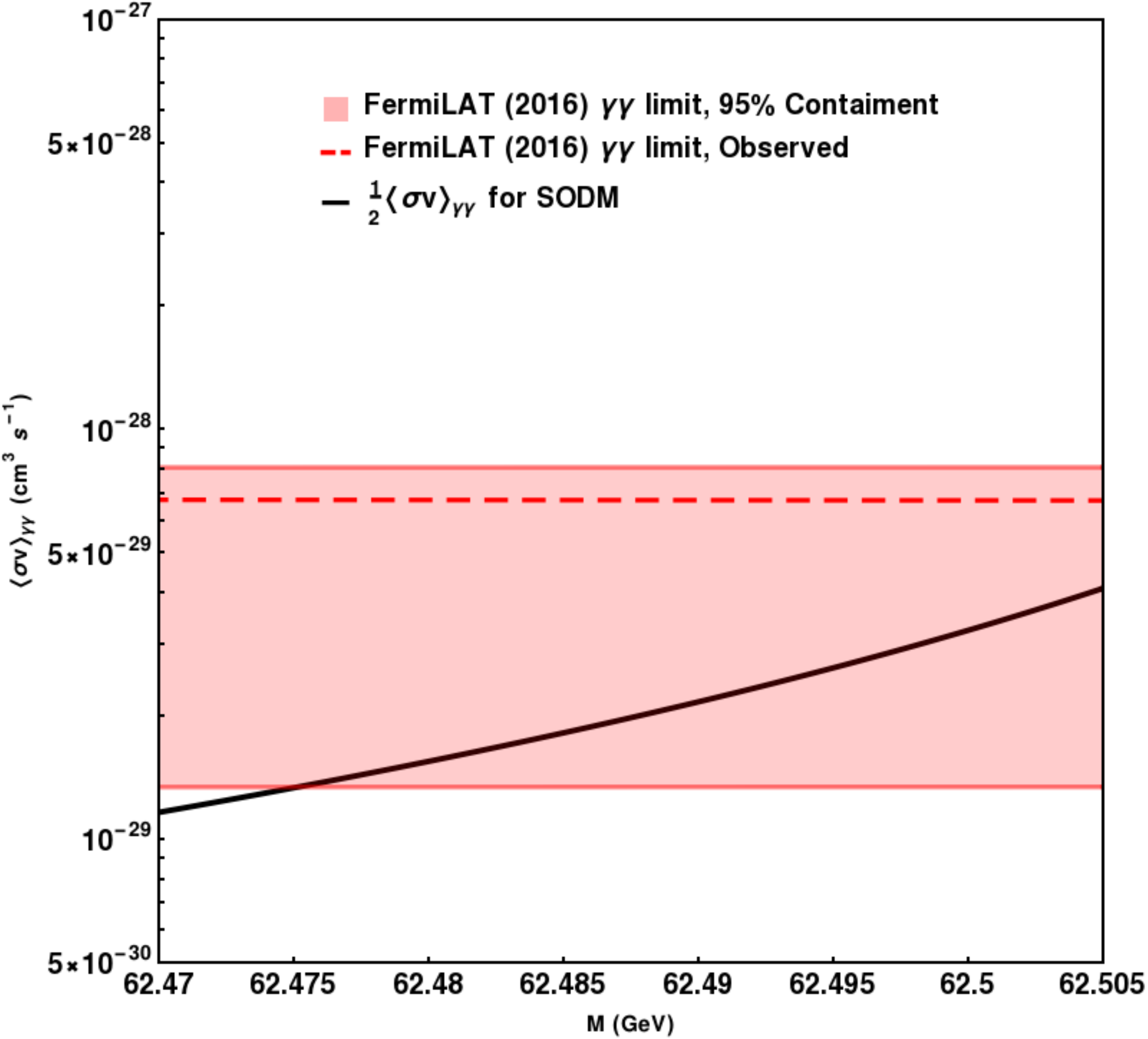}
\caption{Annihilation cross section of SODM into two photons  for  $g_{s}=9.81\times 10^{-4}$ (solid line). The dashed 
line correspond to the upper limits and the shadow band to the $95\%$ confidence level region obtained in \cite{Ackermann:2015lka}.  }
\label{twophotons}
\end{figure}

In Fig. \ref{twophotons} we show the upper bounds obtained by FermiLAT \cite{Ackermann:2015lka} and our results 
for the window of SODM mass consistent with GRE data. In this case the predictions of our formalism are consistent with 
available data but also in this channel we are at edge of the allowed values for  $ \langle \sigma v_{r} \rangle_{\gamma\gamma}$
and lowering this upper bound could test the possibility that dark matter has a $(1,0)\oplus (0,1)$ space-time structure.

 \section{Conclusions}
Summarizing, in this work we study the  production of gamma rays in the annihilation of dark matter with a $(1,0)\oplus (0,1)$ 
space time structure. We start noticing that values for the average cross sections of the annihilation into fermions 
$\langle \sigma v_{r} \rangle_{f}\approx 10^{-26}cm^{3}/seg$ obtained  in the fit of the GRE data done in \cite{Calore:2014nla},
can be obtained for SODM only in the case of resonant processes. Next we perform a systematic study of all the possible mechanisms 
offered by SODM which produce photonic signals. Mechanisms for prompt photons are classified as initial state radiation, internal 
bremsstrahlung and final state radiation. We study also delayed photon emission from the inverse Compton scattering of electrons 
produced directly in the annihilation $\bar{D}D\to e^{+}e^{-}$ or secondary electrons from the subsequent decay or hadronization of 
fermions produced in $\bar{D}D\to \bar{f}f$ with $f=\mu,\tau,q,c,b$. We find that, for SODM, the main contributions come from the 
hadronization of $b$ quarks produced in $\bar{D}D\to \bar{b}b$. There are two mechanisms involved: the prompt photons
from the hadronization products, and the delayed photon emmision in the ICS of electrons produced in the hadronizations 
process.  Similar mechanisms for the $c$ quark, for the $\tau$ lepton decay products and the internal bremsstrahlung, yield 
sub-leading contributions in the low energy 
region but become competitive at the end part of the spectrum. All these mechanisms depend only of the SODM 
mass, $M$, and the scalar Higgs coupling to the SODM, $g_{s}$. Taking into account all these contributions we find a good agreement with 
the GRE data for the windows $g_{s}\in [0.98, 1.01] \times 10^{-3}$ and $M\in [62.470,62.505]~GeV$. 

We check the consistency of these results with the constraints from relic density, direct and indirect detection experiments. Constraints 
on the dark matter-proton cross section $\sigma_{p}$ from 
XENON1T  \cite{Aprile:2017iyp} are well satisfied. Our previous calculation of the relic density is based on the non-relativistic expansion 
which is broken down by the Higgs resonant effects. We perform a new calculation using the full (relativistic) annihilation cross sections 
finding substantial modifications near the resonance. The measured relic density  \cite{Tanabashi:2018oca}, turns out to be consistent 
with the windows for $g_{s}$ and $M$ imposed by the GRE excess data and correlates these parameters according to Fig \ref{gsres}. 
As for indirect detection experiments, we find consistency with constraints on the annihilation cross section of 
$\mu^{+}\mu^{-}$ \cite{Bergstrom:2013jra}, $\tau^{+}\tau^{-}$ \cite{Drlica-Wagner:2015xua} and $\gamma\gamma$ 
\cite{Abdallah:2018qtu},\cite{Ackermann:2015lka}. The consistency with constraints for the annihilation into $\bar{b}b$ 
\cite{Drlica-Wagner:2015xua}, requires a detailed bin-by-bin analysis of the energy flux. We find that also in this case our results 
are consistent with available data.

The results for the annihilation of SODM into $\bar{b}b$ and $\gamma\gamma$ are at the edge of existing upper bounds and lowering 
these limits would give definite tests of the possibility that dark matter has a $(1,0)\oplus (0,1)$ space time structure and a mass 
$M\approx M_{H}/2$, which has been shown here to give a consistent description of the so far calculated observables.

\bibliographystyle{JHEP}
\bibliography{dm}
\end{document}